\documentclass[prl,twocolumn,nofootinbib,superscriptaddress,showpacs,aps,10pt]{revtex4-1}
\usepackage{epsfig}             
\usepackage{epstopdf}           
\usepackage{hyperref}           
\usepackage{color}
\usepackage{colortbl}
\usepackage{graphicx}
\usepackage{amssymb,amsmath}
\usepackage{subfig}
\usepackage{caption}
\usepackage{enumerate}
\usepackage{gensymb}
\usepackage{url}
\usepackage{floatrow}
\usepackage{gensymb}
\usepackage{multirow}
\usepackage{amsmath}
\begin{document}


\title{Spin-parity of the 13.35 MeV state and high-lying states around 20 MeV in excitation energy in $^{12}$C nucleus}

\author{A.S. Demyanova}
\email[Email:~]{a.s.demyanova@bk.ru}
\affiliation{National Research Centre Kurchatov Institute, Akademika Kurchatova pl. 1, 123182 Moscow, Russia}

\author{ V. I. Starastsin}
\affiliation{National Research Centre Kurchatov Institute, Akademika Kurchatova pl. 1, 123182 Moscow, Russia}
\author{A. A. Ogloblin}
\affiliation{National Research Centre Kurchatov Institute, Akademika Kurchatova pl. 1, 123182 Moscow, Russia}
\author{A. N. Danilov}
\affiliation{National Research Centre Kurchatov Institute, Akademika Kurchatova pl. 1, 123182 Moscow, Russia}
\author{S. V. Dmitriev}
\affiliation{National Research Centre Kurchatov Institute, Akademika Kurchatova pl. 1, 123182 Moscow, Russia}

\author{W. H. Trzaska}
\affiliation{Department of Physics, University of Jyv\"askyl\"a, FI-40014 Jyvaskyla, Finland}
\author{P. Heikkinen}
\affiliation{Department of Physics, University of Jyv\"askyl\"a, FI-40014 Jyvaskyla, Finland}

\author{T. L. Belyaeva}
\affiliation{Universidad Autonoma del Estado de Mexico, 50000 Toluca, Mexico}

\author{S. A. Goncharov}
\affiliation{M.V. Lomonosov Moscow State University, Leninskie Gory, 119991 Moscow, Russia}

\author{V. A. Maslov}
\affiliation{Flerov Laboratory for Nuclear Research, JINR, Dubna, Moscow Region 141980, Russia}
\author{Yu. G. Sobolev}
\affiliation{Flerov Laboratory for Nuclear Research, JINR, Dubna, Moscow Region 141980, Russia}

\author{Yu. B. Gurov}
\affiliation{National Research Nuclear University MEPhI, Kashirskoe sch. 31, 115409 Moscow, Russia}

\author{N. Burtebaev}
\affiliation{Institute of Nuclear Physics, National Nuclear Center of Republic of Kazakhstan, Almaty 050032, Republic of Kazakhstan}

\author{ D. Janseitov}
\affiliation{Institute of Nuclear Physics, National Nuclear Center of Republic of Kazakhstan, Almaty 050032, Republic of Kazakhstan}
\affiliation{Kazakh National University, al-Farabi 71, 050040 Almaty, Kazakhstan}

\date{\today}

\begin{abstract}
Study of the $^{11}$B($^{3}$He,d)$^{12}$C reaction at incident $^{3}$He energy E$_{lab}$ = 25 MeV has been performed at the K-130 cyclotron at the University of Jyv\"askyl\"a, Finland. Differential cross sections have been measured for the 13.35 MeV state and for the states with excitation energy around 20 MeV. The data were analyzed with the DWBA method. A tentative assignment, 4$^{-}$, is given for the state at 13.35 MeV. For the state at 20.98 MeV, the spin-parity 3$^{-}$ and the isospin T = 0 are assigned for the first time. Our model description of the broad state at 21.6 MeV is consistent with the previous assignments of isospin T = 0 and spin-parity of 2$^{+}$ and 3$^{-}$. The excited state at 22.4 MeV may have possible spin-parities of either 6$^{+}$ or 5$^{-}$. The collected statistics was insufficient to solve this question.

\end{abstract}

\keywords{$^{11}$B($^{3}$He,d)$^{12}$C reaction; high-lying excited states of $^{12}$C; spin-parities; rotational bands; nuclear radii; DWBA analysis.}

\maketitle


\section{Introduction}

Recently $^{12}$C become the key nucleus in the study and description of nucleon clustering in light nuclei. For instance, identification of the states with abnormally large radii, observation of alpha-cluster rotational bands, and the prevailing absence of adequate understanding of the structure of the famous Hoyle state (0$^{+}_{2}$, $E_{x}$ = 7.65 MeV), evoked numerous theoretical and experimental studies (as example Ref. \cite{M.Kamimura_1981, Riisager_1992, Otsuka_1993, Hansen_1995, W_von_Oertzen, Chernykh_2007, Demyanova_2008, 7, 10, 11, Tanihata_2013, 4}).

The structure of the 0$^{+}_{2}$, 7.65 MeV Hoyle state of $^{12}$C permanently attracts attention due to its importance in understanding clustering in nuclei and the key role in nucleosynthesis in the Universe. During last decade there appeared several new theoretical approaches that predicted some unusual features of this state. The most ambitious among them is the model of $\alpha$-particle condensation (APC)\cite{1, 2, 3} according to which the Hoyle state is resembled as a gas of almost noninteracting alpha particles, and $^{12}$C in this state is expected to have anomalously enhanced radius, 60-80$\%$ larger than that in the ground state (g.s.). A number of different theoretical calculations also predict an enlarged radius of $^{12}$C in the Hoyle state, but smaller than by APC (see, for example, Refs.\cite{4, 5} and references therein). 

A direct experimental method, the Modified Diffraction Model (MDM)\cite{6}, applied to the analysis of inelastic scattering of various nuclei on $^{12}$C, gave an increase of 25$\%$ of the $rms$ radius (2.89 $\pm$ 0.04 fm)\cite{6} of $^{12}$C in the Hoyle state comparing with that in the ground state. A similar result was obtained by the Antisymmetrized Molecular Dynamics (AMD) calculations\cite{7}.

A special interest is generated by the proposed existence of the excited states \textquotedblleft genetically related\textquotedblright to the Hoyle state. The idea that the Hoyle state might be the head of a rotational band became quite natural after appearance of the Morinaga's model\cite{8} presenting this level as a chain-like configuration of three $\alpha$-particles. Recently, a suitable candidate for the second member of the rotational band based on the Hoyle state was identified either at 9.75 or at 10.13 MeV \cite{9, 10, 11}. The radius of this state was determined to be equal $\approx$ 3.1 fm\cite{12}, i.e. practically the same as that for the Hoyle state. Thus this state can be regarded as a second member of the rotational band based on the Hoyle state. 

Apparently, the members of the rotational band based on the Hoyle state are not the only states of $^{12}$C with an enlarged radius. A considerable size increase was found also for other states located above the $^{12}$C $\rightarrow$ 3$\alpha$ threshold. The radius of the 3$^{-}$ state at 9.64 MeV was determined to be $R$ = 2.88$\pm$0.11 fm\cite{6}. The exotic 3$\alpha$ and $\alpha$ + $^{8}$Be structures of the states in $^{12}$C near and above the $\alpha$ - emission threshold remain objects of intense theoretical studies\cite{5}. They include calculations of the $^{8}$Be direct transfer\cite{13} contributing to the elastic and inelastic $\alpha$ + $^{12}$C scattering to the 2$^{+}_{1}$ , 0$^{+}_{2}$ and 3$^{-}_{1}$ states. The measurements were done at 110 MeV\cite{14, 23} over the full angular range. 

The relative contributions of different angular momenta of $\alpha$ and $^{8}$Be in the four lowest states of $^{12}$C were calculated as a ratio of the extracted spectroscopic factors corresponding to the angular momenta $L$ = 0, 2, and 4 for the 0$^{+}_{1,2}$ and 2$^{+}_{1}$ states, and $L$ = 1, 3 and 5 for the 3$^{-}_{1}$ state. A comparison of these ratios revealed interesting regularities. Namely, the occupation probability in the g.s. and the 2$^{+}$ 4.44-MeV state (members of the g.s. rotational band) was found almost evenly distributed between all orbital momenta. The occupation probabilities in the 0$^{+}_{2}$ Hoyle state and the 3$^{-}$ 9.64-MeV state (the first members of the positive and negative rotational bands) were found predominantly concentrated in the lowest orbit with $L$ = 0 and 1, and probabilities 62$\%$ and 69$\%$, respectively. This fact indicates that the structure of the 0$^{+}_{2}$ and the 3$^{-}$ 9.64-MeV states is completely different from the structure of lowest $^{12}$C states. 

There are few open questions regarding excited states of $^{12}$C and the rotational bands that can exist in this nucleus. The g.s. rotational band (0$^{+}$, g.s.; 2$^{+}$, 4.44 MeV; 4$^{+}$, 14.08 MeV) in $^{12}$C is well known. Recently, the authors of Ref.\cite{15}, based on the D$_{3h}$ symmetry of $^{12}$C, suggested that a negative branch of the g.s. rotational band in $^{12}$C includes the states: 3$^{-}$, 9.64 MeV; 4$^{-}$, 13.35 MeV and the new high spin $J^{\pi}$ = 5$^{-}$ state at 22.4(2) MeV in $^{12}$C. Unfortunately, the spin-parity of the 13.35 MeV state remains ambiguous with contradicting assignments of either 2$^{-}$ \cite{16,17} or 4$^{-}$ \cite{18,19}. Resolving this ambiguity is important for the understanding the structure of $^{12}$C as a whole. 

No less interesting is the situation with the rotational band based on the Hoyle state.

While the 2$^{+}_{2}$ state at 9.75 or 10.13 MeV\cite{9, 10, 11} is quite certain today, the 4$^{+}_{2}$ state has raised many questions, in particular, related to the part of the spectrum near 13-14 MeV. Recently we have identified a new level in $^{12}$C at 13.75 $\pm$ 0.12 MeV ($\Gamma$ = 1.4 $\pm$ 0.15 MeV)\cite{20} with a spin-parity assignment of 4$^{+}$. Apparently, this state coincides with the state at 13.3 $\pm$ 0.2 MeV ($\Gamma$ = 1.7 $\pm$ 0.2 MeV) in $^{12}$C previously determined in Ref.\cite{21}. If so, the state at 13.75 MeV can be included into the Hoyle-state based rotational band as a third member\cite{4, 5}. 

The next challenge is identification of spin-parities of all the states with excitation energy between 18 and 23 MeV. In this region, one would expect to find the higher members of the $^{12}$C rotational bands. 

To address these questions, we have studied the $^{11}$B($^{3}$He,d)$^{12}$C reaction at the incident $^{3}$He energy of the 25 MeV.  Accordingly to our estimates, this energy is optimal for a spin-parity determination of the higher-excited levels in the 18-23 MeV region. For the 13.35-MeV state, the shape of the angular distributions in the c.m. angular range 4$^{\circ}$ - 60$^{\circ}$ is quite different for the transferred angular momenta $L$ = 0 and $L$ = 2. This should allow for an unambiguous determination of spin-parity for this state. The available published data on the transfer reaction $^{11}$B($^{3}$He,d)$^{12}$C at $E$($^{3}$He) = 44 MeV\cite{17} are too scant (the angular distribution contains only 3 points) to allow for a reliable determination of the spin-parity of the 13.35-MeV state.

\section{Experimental procedure}

The measurements were carried out in the 150 cm diameter Large Scattering Chamber (LSC)\cite{22} at the Accelerator Laboratory of the University of Jyv\"askyl\"a (Finland). The $^{3}$He beam at $E$($^{3}$He) = 25 MeV was extracted from the K-130 cyclotron. The LSC was equipped with three sets of $\Delta E$ - $E$ detector telescopes, each containing two independent $\Delta E$ detectors and one common $E$ detector. So each device allowed carrying out measurements at two angles. The measurements in c.m. angular range 10$^{\circ}$ were conducted in one exposure. Two silicon pin diodes of 380 and 100 $\mu$m were operating as $\Delta E$ detectors and 3.6 mm lithium-drifted silicon detectors as $E$ detectors. The differential cross sections of the $^{11}$B($^{3}$He,d)$^{12}$C reaction were measured over the 4$^{\circ}$ - 60$^{\circ}$ range in c.m. The beam intensity was about 20 particle nA. A self-supported, 0.275 mg/cm$^{2}$ thick, enrichment (95$\%$) boron foil was used as a $^{11}$B target. 

The $^{10}$B nuclei were the only impurity in the target. Presence of the lighter isotop added excited states of $^{11}$C to the experimental spectra. Fortunately, this did not compromise our measurements as all the levels below the excitation energy of 19 MeV were well separated. This was possible thanks to the total energy resolution of about 80$-$120 keV. 

This very good energy resolution was needed to resolve the 4$^{-}$ or 2$^{-}$ and 1$^{+}$ states in $^{12}$C. It was achieved with a monochromatization method described in Ref.\cite{22}. The procedure reduces the energy spread of the native cyclotron beam by a factor of 2 to 3 making this measurement possible. Fig. 1 shows a sample of the registered deuteron spectrum from the reaction $^{11}$B($^{3}$He,d)$^{12}$C.

Peaks corresponding to the relevant transitions in the collected deuteron spectra were identified and parametrized using a standard Gaussian decomposition method. With the known energy calibration, the peak positions and widths were fixed in accordance with the generally accepted values while the areas under the peaks were treated as free parameters. At excitation energies above 19 MeV (Fig. 1 b) the procedure became more complex as some of the states could not be fully resolved. 

In the ($^{3}$He,d) reaction, both the isospin $T$ = 0 and $T$ = 1 states are excited while in the inelastic scattering of $\alpha$-particle, only the states with $T$ = 0 are excited. Therefore, if one observes the same state in the ($^{3}$He,d) reaction and in the inelastic scattering spectrum, one can confidently assign $T$ = 0 to this state. This comparative procedure was applied to the experimental data on the inelastic scattering of $\alpha$-particles on $^{12}$C at $E_{lab}$ = 110 MeV \cite{14, 23}. In Fig 2, we show a sample spectrum from the $^{4}$He + $^{12}$C scattering at $\theta_{lab}$ = 23$^{\circ}$ showing the excitation of $^{12}$C states around 21 MeV.

\begin{figure} [th]
\centerline{\includegraphics[width=9 cm]{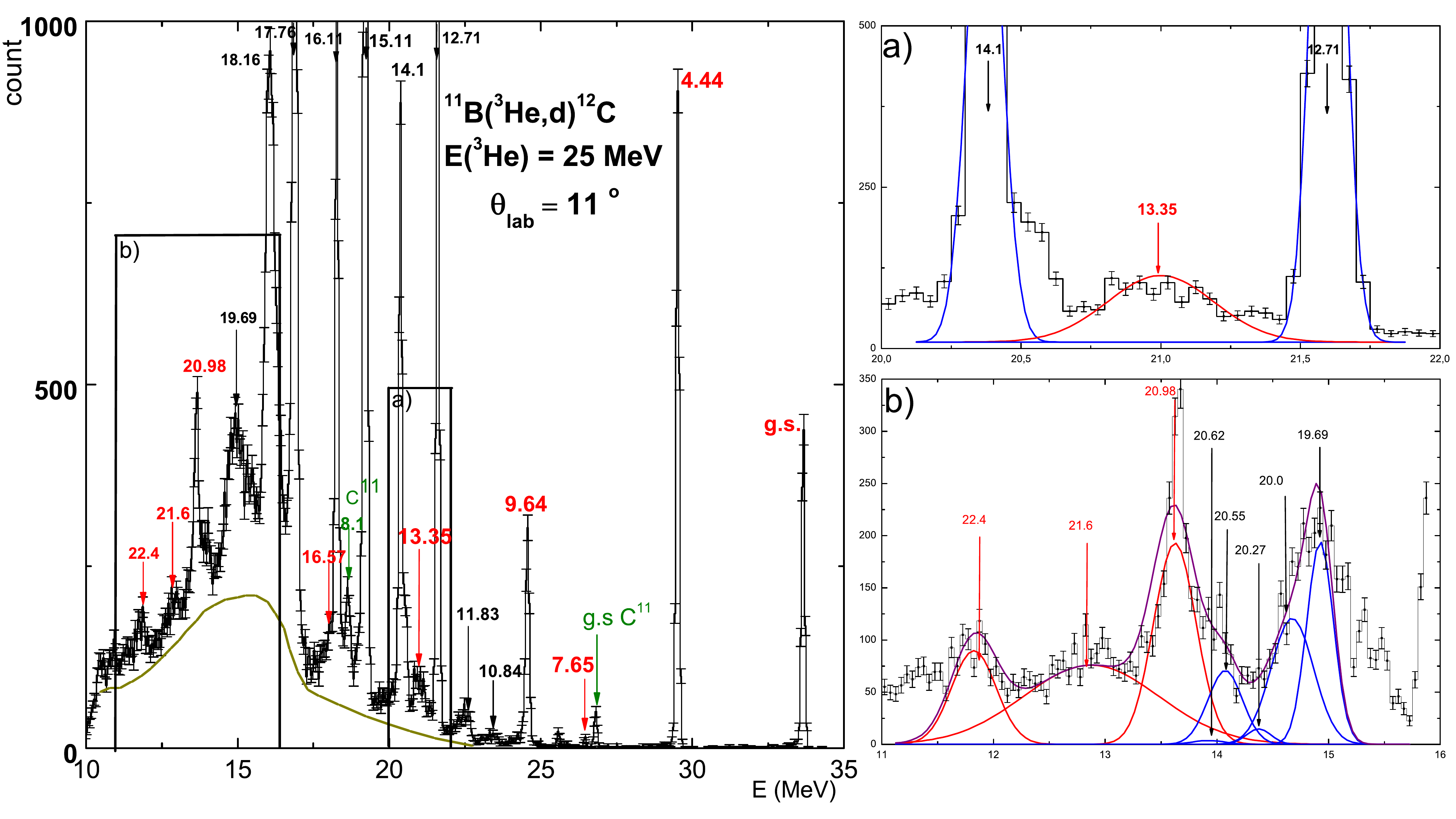}}
\caption{
Left panel: sample of a deuteron spectrum from the $^{11}$B($^{3}$He,d)$^{12}$C reaction at $E_{lab}$ = 25 MeV registered at $\theta_{lab}$= 11$^{\circ}$. The solid dark green curve corresponds to the background. The black rectangular areas labeled a) and b) are shown expanded in the right panels. Right panel a): region near 13.35 MeV in excitation energy of $^{12}$C.
Right panel b): region near 21 MeV in excitation energy of $^{12}$C. Red color indicates states that are of particular interest to us. Blue color indicates other states.
}
\end{figure}

\begin{figure} [th]
\centerline{\includegraphics[width=9 cm]{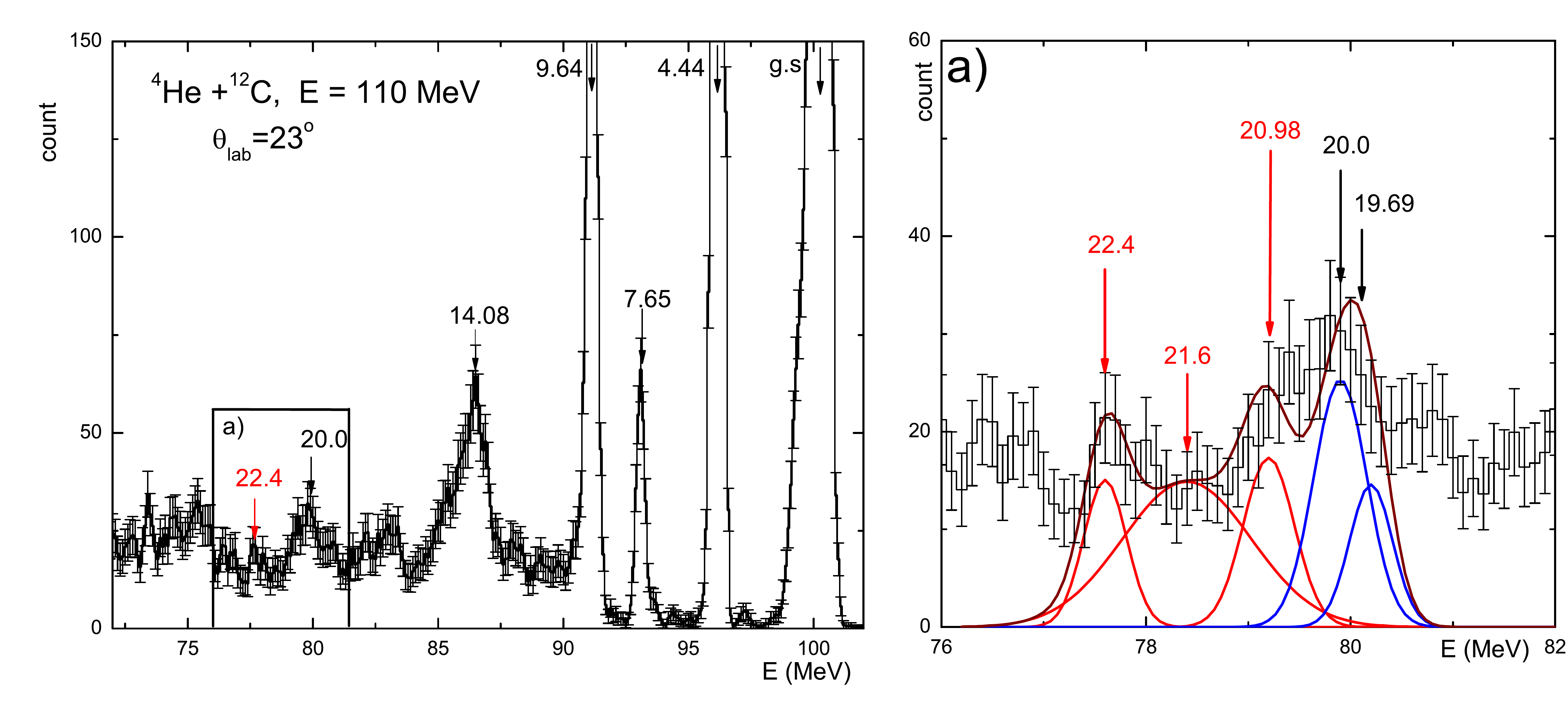}}
\caption{Left panel: a sample spectrum from the $^{4}$He + $^{12}$C scattering at $E_{lab}$=110 MeV for $\theta_{lab}$=23$^{\circ}$. Right panel: an expanded part of the spectrum region near 21 MeV. Red color indicates the states that are of particular interest to us. Blue color indicates other states.
}
\end{figure}

In total, deuteron angular distribution for the g.s. and eight excited states of $^{12}$C were extracted using the $^{11}$B($^{3}$He,d)$^{12}$C. The resulting differential cross sections for the g.s. and the excited states at $E_{x}$ = 4.44, 7.65, 9.64, 13.35, 16.57, 20.98, 21.6 and 22.4 MeV are presented in Figs. 3-7 and 10. 

\section{Theoretical analysis}

Theoretical analysis of the experimental differential cross sections are carried out in the framework of the finite range distorted-wave Born approximation (DWBA)\cite{24} using the FRESCO code\cite{25}. The contributions of all allowed combinations of transmitted angular moments $L$ and spins $J$ are coherently accounted.

The elastic scattering wave functions in the entrance ($^{3}$He + $^{11}$B) and the exit (d + $^{12}$C) reaction channels are calculated in the framework of the fenomenological optical model. A standard Woods-Saxon form of the real part of optical potentials and a combination of the volume $W_{s}$ and surface $W_{d}$ potentials for the imaginary part are used. For the entrance channel, we have applied parameters of the global potential\cite{26}. For the exit channel, also global potential parameters were used\cite{27}. These parameterizations give a good description of the elastic scattering data in the forward angular range (up to 40$^{\circ}$). 

The transfer form factors to the bound states are modeled by the normalized single-particle wave functions in the Woods-Saxon potentials. The depth of the potential is automatically varied to give a binding energy of the transmitted particle (Well-Depth-Prescription procedure, WDP). The relative contributions ($A_{LJ}$) of components corresponding to different moments ($LJ$) and the geometric parameters $r_{0}$ and $a$ of the Woods-Saxon potential are adjusted to fit the experimental angular distributions especially in the main front peak range. The geometrical parameters and the relative contribution of components $A_{LJ}$ for each bound state of the final nucleus are presented In Table 1.

\begin{table}[h]
\caption{Geometrical parameters of the single-particle potential of p + $^{11}$B in $^{12}$C and the relative contribution $A_{LJ}$ of components for allowed combinations of angular momentum transfers $LJ$.}
{\begin{tabular}{@{}cccccc@{}} \toprule
$^{12}$C states & $r_{0}$ & $a$ & $L$ & $J$ & $A_{LJ}$
\\
& (fm) & (fm) \\ \colrule
g.s, 0$^{+}$\hphantom{00} & \hphantom{0}1.1 & \hphantom{0}0.87 & \hphantom{0}1 & \hphantom{0}3/2 & \hphantom{0}1.00\\
\\
4.44, 2$^{+}$\hphantom{00} & \hphantom{0}1.0 & \hphantom{0}1.0 & \hphantom{0}1 & \hphantom{0}1/2 & \hphantom{0}1.00\\
\\
7.65, 0$^{+}$\hphantom{00} & \hphantom{0}0.6 & \hphantom{0}0.1 & \hphantom{0}1 & \hphantom{0}3/2 & \hphantom{0}1.00\\
\\
9.64, 3$^{-}$\hphantom{00} & \hphantom{0}1.2 & \hphantom{0}1.5 & \hphantom{0}2 & \hphantom{0}3/2 & \hphantom{0}1.00\\
&&& \hphantom{0}2 & \hphantom{0}5/2 & \hphantom{0}0.645\\
\\
13.35, 2$^{-}$\hphantom{00} & \hphantom{0}1.1 & \hphantom{0}1.3 & \hphantom{0}0 & \hphantom{0}1/2 & \hphantom{0}0.197\\
&&& \hphantom{0}2 & \hphantom{0}3/2 & \hphantom{0}0.661\\
&&& \hphantom{0}2 & \hphantom{0}5/2 & \hphantom{0}0.661\\
&&& \hphantom{0}4 & \hphantom{0}7/2 & \hphantom{0}0.296\\
\\
13.35, 4$^{-}$\hphantom{00} & \hphantom{0}1.2 & \hphantom{0}1.4 & \hphantom{0}2 & \hphantom{0}5/2 & \hphantom{0}0.79\\
&&& \hphantom{0}4 & \hphantom{0}7/2 & \hphantom{0}0.118\\
&&& \hphantom{0}4 & \hphantom{0}9/2 & \hphantom{0}0.118\\
&&& \hphantom{0}6 & \hphantom{0}11/2 & \hphantom{0}0.59\\
\\ \botrule
\end{tabular}}
\end{table}

\newpage

The proton transfer form factor to the continuum states of $^{12}$C is calculated using the wave functions $\Phi (r)$ defined in Ref.\cite{25}
\begin{equation}
    \label{Eq1}
    \Phi(r) = \sqrt{\frac{2}{\pi N}}\int_{k_{1}}^{k_{2}}w(k)\varphi_{k, LJ}(r)dk
\end{equation}

where $N$ is the normalization of the weight function 
\begin{equation}
    \label{Eq2}
    w(k)=exp(-i\delta_{k})\sin{\delta_{k}}
\end{equation}

The wave functions $\Phi(r)$ are normalized to unity in the sufficiently large interval [0, $R$]. In our case, the maximum radius $R$ = 150 fm was sufficient, so a single-particle wave function $\varphi_{k}(r)$ is averaged over energy “bin” in continuum\cite{25}. The experimental level widths\cite{16} were used as the resonance widths. The relative contribution of components $A_{LJ}$, potential depths $V$ and geometric parameters $r_{0}$ and $a$ of Woods-Saxon potential selected for the best fit of the experimental angular distributions at forward angles are presented in Table 2 for each state of the continuum proton spectrum of $^{12}$C.

\begin{table}[h]
\caption{The potential parameters ($V$, $r_{0}$ and $a$) of the single-particle potential of p+$^{11}$B for the continuum states of $^{12}$C, relative contribution $A_{LJ}$ of components for allowed combinations of angular momentum transfers $LJ$.}
{\begin{tabular}{@{}ccccccc@{}} \toprule
$^{12}$C states & $V$ & $r_{0}$ & $a$ & $L$ & $J$ & $A_{LJ}$
\\
& (MeV) & (fm) & (fm) \\ \colrule
16.57, 2$^{-}$\hphantom{00} & \hphantom{0}160 & \hphantom{0}1.2 & \hphantom{0}0.7 & \hphantom{0}0 & \hphantom{0}1/2 & \hphantom{0}0.474\\
&&&& \hphantom{0}2 & \hphantom{0}3/2 & \hphantom{0}0.623\\
&&&& \hphantom{0}2 & \hphantom{0}5/2 & \hphantom{0}0.623\\
\\
20.98, 3$^{-}$\hphantom{00} & \hphantom{0}175 & \hphantom{0}1.7 & \hphantom{0}0.8 & \hphantom{0}2 & \hphantom{0}3/2 & \hphantom{0}0.639\\
&&&& \hphantom{0}2 & \hphantom{0}5/2 & \hphantom{0}0.639\\
&&&& \hphantom{0}4 & \hphantom{0}7/2 & \hphantom{0}0.303\\
&&&& \hphantom{0}4 & \hphantom{0}9/2 & \hphantom{0}0.303\\
\\
20.98, 5$^{-}$\hphantom{00} & \hphantom{0}161 & \hphantom{0}1.6 & \hphantom{0}0.9 & \hphantom{0}4 & \hphantom{0}7/2 & \hphantom{0}0.5\\
&&&& \hphantom{0}4 & \hphantom{0}7/2 & \hphantom{0}0.5\\
&&&& \hphantom{0}6 & \hphantom{0}7/2 & \hphantom{0}0.5\\
&&&& \hphantom{0}6 & \hphantom{0}7/2 & \hphantom{0}0.5\\
\\
21.6, 2$^{+}$\hphantom{00} & \hphantom{0}180 & \hphantom{0}1.2 & \hphantom{0}1.2 & \hphantom{0}1 & \hphantom{0}1/2 & \hphantom{0}0.5\\
&&&& \hphantom{0}1 & \hphantom{0}3/2 & \hphantom{0}0.5\\
&&&& \hphantom{0}1 & \hphantom{0}5/2 & \hphantom{0}0.5\\
&&&& \hphantom{0}1 & \hphantom{0}7/2 & \hphantom{0}0.5\\
\\
21.6, 3$^{-}$\hphantom{00} & \hphantom{0}185 & \hphantom{0}1.2 & \hphantom{0}1.2 & \hphantom{0}2 & \hphantom{0}3/2 & \hphantom{0}0.697\\
&&&& \hphantom{0}2 & \hphantom{0}5/2 & \hphantom{0}0.697\\
&&&& \hphantom{0}4 & \hphantom{0}7/2 & \hphantom{0}0.116\\
&&&& \hphantom{0}4 & \hphantom{0}9/2 & \hphantom{0}0.116\\
\\
22.4, 5$^{-}$\hphantom{00} & \hphantom{0}155 & \hphantom{0}1.8 & \hphantom{0}0.7 & \hphantom{0}4 & \hphantom{0}7/2 & \hphantom{0}0.588\\
&&&& \hphantom{0}4 & \hphantom{0}9/2 & \hphantom{0}0.588\\
&&&& \hphantom{0}6 & \hphantom{0}11/2 & \hphantom{0}0.425\\
&&&& \hphantom{0}6 & \hphantom{0}13/2 & \hphantom{0}0.425\\
\\
22.4, 6$^{+}$\hphantom{00} & \hphantom{0}150 & \hphantom{0}1.8 & \hphantom{0}0.8 & \hphantom{0}5 & \hphantom{0}9/2 & \hphantom{0}0.5\\
&&&& \hphantom{0}5 & \hphantom{0}11/2 & \hphantom{0}0.5\\
&&&& \hphantom{0}7 & \hphantom{0}13/2 & \hphantom{0}0.5\\
&&&& \hphantom{0}7 & \hphantom{0}15/2 & \hphantom{0}0.5\\
\\ \botrule
\end{tabular}}
\end{table}

\newpage

\section{Results and discussion}

Figure 3 shows the experimental deuteron angular distributions in comparison with the DWBA calculations for the g.s (Fig 3a) and the states at 4.44 MeV (Fig. 3b), 7.65 MeV (Fig. 3c), and 9.64 MeV (Fig. 3d). For these levels, the values of spin-parity and isospin are well known\cite{16}. The calculations satisfactorily describe the data, especially at forward angles, and allow us to correctly determine the spin-parity values and the transfer moments (see Table 1).

\begin{figure} [th]
\centerline{\includegraphics[width=8 cm]{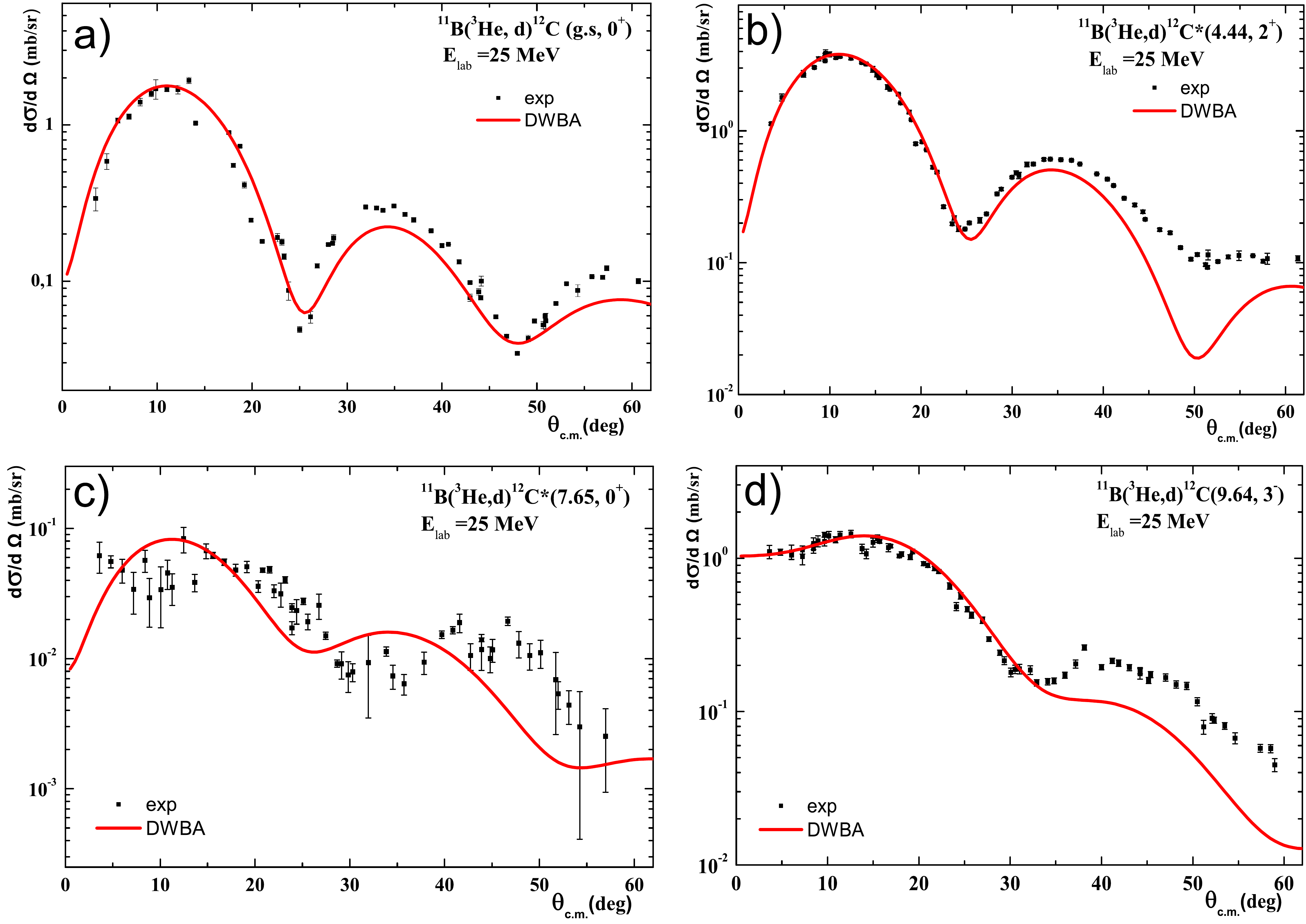}}
\caption{The experimental differential cross sections of the $^{11}$B($^{3}$He,d)$^{12}$C reaction at $E_{lab}$ = 25 MeV for (a) ground state, (b) 4.44-MeV state, (c) 7.65-MeV state, and (d) 9.64-MeV state. The red solid curves show DWBA calculations for these states.
}
\end{figure}

\newpage

In the following chapters we summarize the outcome of the analysis of studying the differential cross sections for the $^{12}$C states at 16.57 MeV, 21.6 MeV, 20.98 MeV, 13.35 MeV, and 22.4 MeV excitation energy.

\subsection{The state at 16.57 MeV and 21.6 MeV}

The differential cross section of the $^{11}$B($^{3}$He,d)$^{12}$C reaction with excitation of the 16.57-MeV state is presented in Fig. 4 in comparison with the DWBA calculation assigning to this state spin-parity of 2$^{-}$ (solid red curve).  As it was mentioned earlier, this excited state was not observed in the inelastic scattering of $\alpha$-particles on $^{12}$C at 110 MeV\cite{14,23} because of the accepted isospin of $T$ = 1\cite{16}.

\begin{figure} [th]
\centerline{\includegraphics[width=6 cm]{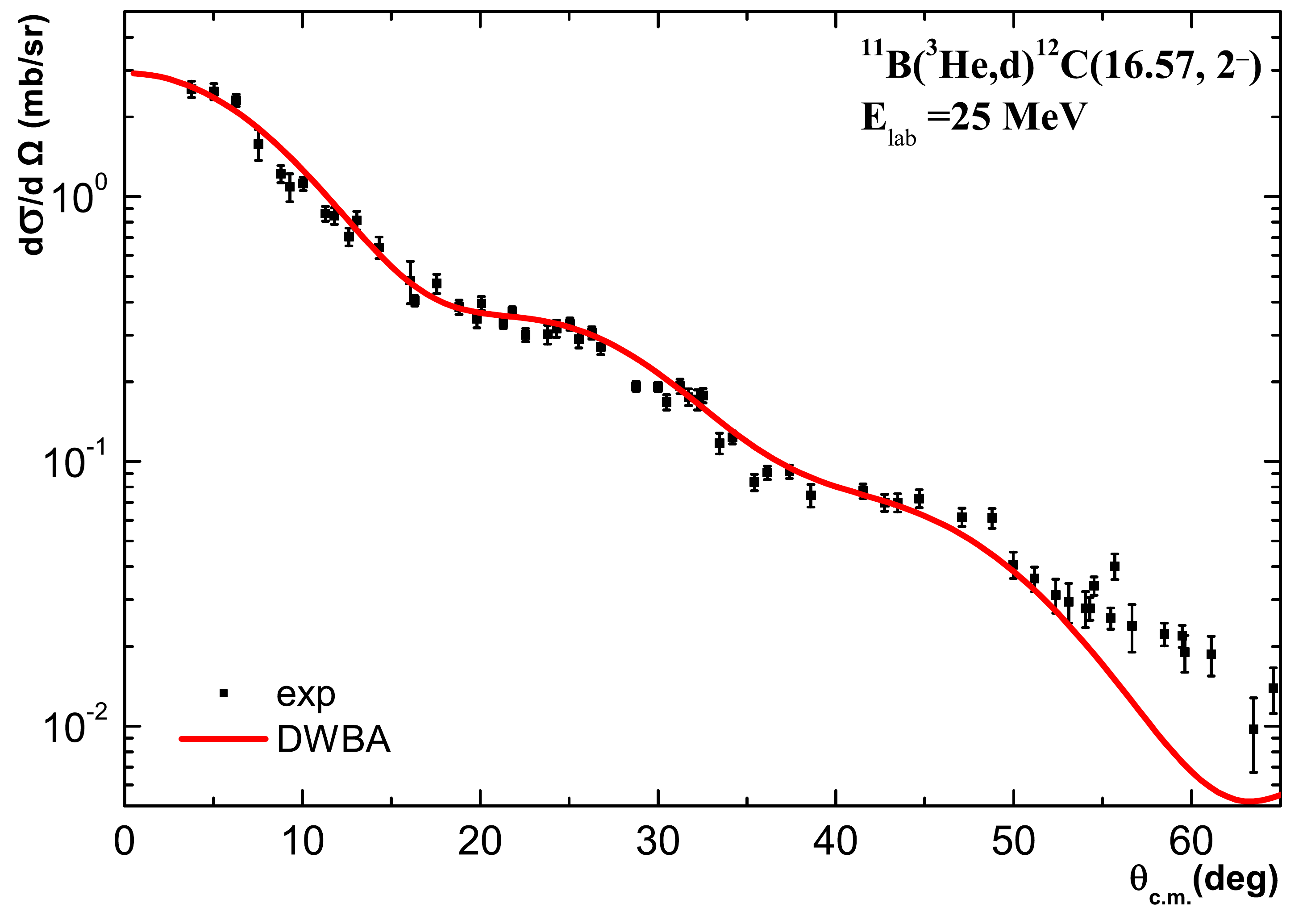}}
\caption{
The experimental differential cross section of the $^{11}$B($^{3}$He,d)$^{12}$C reaction at $E_{lab}$ = 25 MeV with the excitation of the 16.57-MeV state (black dots). The red solid curve corresponds to the DWBA calculation for this state with spin-parity 2$^{-}$.
}
\end{figure}

\newpage 

The experimental differential cross section of the $^{11}$B($^{3}$He,d)$^{12}$C reaction with an excitation of the 21.6-MeV state is plotted in Fig. 5 in comparison with the DWBA calculations assuming generally accepted  2$^{+}$ and 3$^{-}$ values of spin-parity\cite{16} for this state. In the right panels of Fig. 5, the DWBA calculations are shown in a linear scale for spin-parity of 3$^{-}$ (green curve in the top panel) and 2$^{+}$ (blue curve in the bottom panel). The behavior of both curves is very similar over the entire angular range. Incoherent sum of the calculated cross sections corresponding to the spin-parities of 2$^{+}$ and 3$^{-}$ with equal weights of 0.5 is shown in Fig. 5 by the red curve to better fit the data. 

\begin{figure} [th]
\centerline{\includegraphics[width=9 cm]{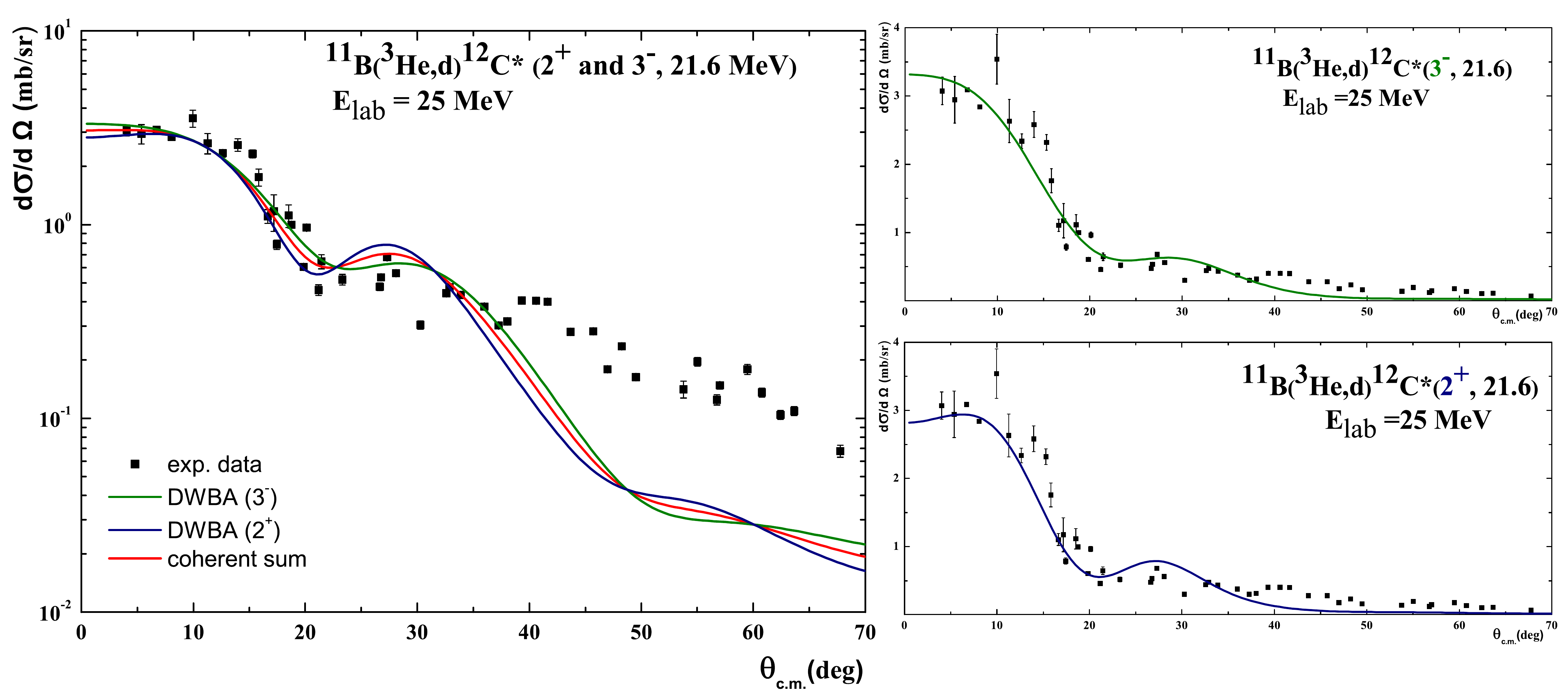}}
\caption{
Left panel: the experimental differential cross section of the $^{11}$B($^{3}$He,d)$^{12}$C reaction at $E_{lab}$ = 25 MeV for the 21.6-MeV state (black dots) of $^{12}$C.  The green and blue solid curves show DWBA calculations with 3$^{-}$ and 2$^{+}$ spin-parities, respectively. Their coherent sum is indicated by a red solid curve. Right panel: the differential cross sections are presented in a linear scale. Top and bottom panels present the data for 3$^{-}$ and 2$^{+}$ spin-parities, respectively.
}
\end{figure}

As the 21.6-MeV state was also observed in the inelastic scattering of $\alpha$-particles at 110 MeV\cite{14,23}, we confirm the generally accepted values of spin-parity of 2$^{+}$ and 3$^{-}$ and the isospin of $T$ = 0 for the 21.6-MeV state of $^{12}$C.

\subsection{The state at 20.98 MeV}

The spin-parity and the isospin have not been so far assigned for the state at 20.98 MeV. The differential cross sections of the $^{11}$B($^{3}$He,d)$^{12}$C reaction with excitation of the 20.98-MeV state are presented in Fig. 6. The DWBA calculations show that the best fit of the data is achieved by choosing the spin-parity of 3$^{-}$. Since the state at 20.98 MeV is observed in the inelastic scattering, we can assign to it the isospin of $T$ = 0.

\begin{figure} [th]
\centerline{\includegraphics[width=6 cm]{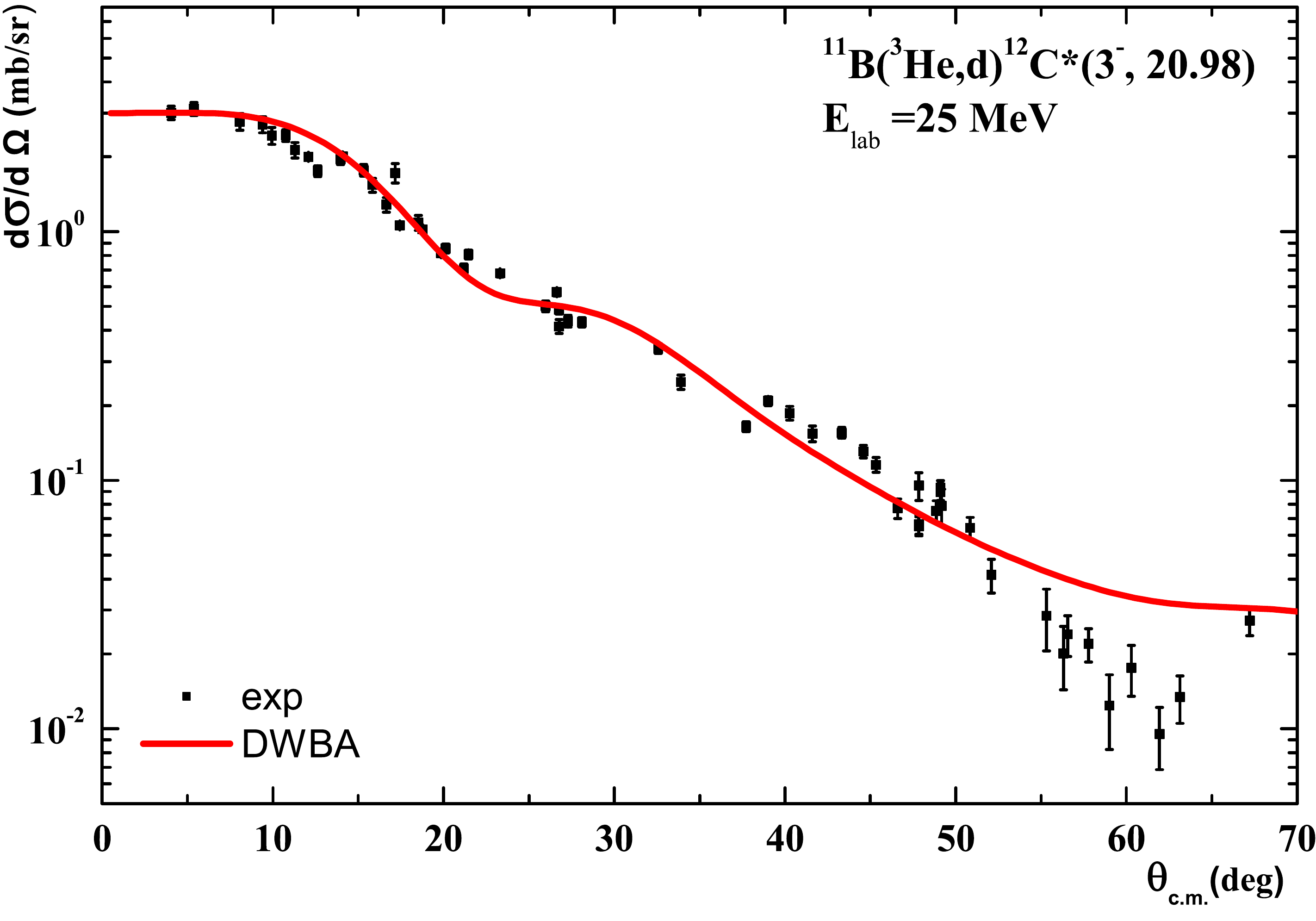}}
\caption{The experimental differential cross section of the $^{11}$B($^{3}$He,d)$^{12}$C reaction to the 20.98-MeV state at $E_{lab}$ = 25 MeV (black dots). The red solid curve corresponds to the DWBA calculations with the spin-parity of 3$^{-}$.
}
\end{figure}

Thus, for the first time, we determine the possible values of spin-parity $J^{\pi}$ = 3$^{-}$ and isospin $T$ = 0 for the state at 20.98 MeV.

\subsection{The state at 13.35 MeV}

The differential cross section of the $^{11}$B($^{3}$He,d)$^{12}$C reaction with excitation of the 13.35 state is presented in Fig. 7. The DWBA analysis reveals that the dominant transferred angular momentum is $L$ = 2. For this moment, there are two possible values of spin-parity: 4$^{-}$ and 2$^{-}$. The DWBA calculations depicted in Fig. 7 with spin-parities of 4$^{-}$ and 2$^{-}$ (red and blue curves, respectively) differ little in the description of the data. In this case, a choice of the spin-parity value of the 13.35-MeV state is hindered and the question remains open.


To answer the question we tried to use our data on the inelastic scattering angular distribution of $\alpha$-particles at 110 MeV\cite{14,23}. 

Since the assumed values of spin-parity are unnatural, one should consider two-step mechanism to excite this state in inelastic scattering. But using this approach we could not reproduce the experimental angular distributions, and moreover the absolute values of the calculated cross sections have been very small. We use an alternative way to estimate the relative contribution of various transferred angular momenta $L$ in the framework of one-step DWBA applying a simple cluster model of $^{12}$C. In this way the inelastic form factor is considered in the frame of microscopic (cluster) interaction model

\begin{multline}
    f_{L}(r)=4\pi V_{0}\sqrt{(2J_{A}+1)}\langle j^{\prime} I_{c} J_{B} \|Y_{L}\| j I_{c} J_{A}\rangle \\ \times \int dr R_{l^{\prime}I^{\prime}j^{\prime}}(r) R_{lIj}(r)
\end{multline}

where $R_{lIj}(r)$ are the single-particle wave functions describing the relative motion (with orbital momentum $l$ and total momentum $j$) of $^{8}$Be (with spin $I$) in the potential of alpha-particle core (with spin $I_{c}$). These wave functions are calculated with a standard WDP procedure using a Woods-Saxon potential. The geometric parameters of the Woods-Saxon potential are selected based on the best fit of the experimental angular distribution. The Gaussian form for a radial part of the central interaction $\nu_{L}$ is used with inverse range parameter $\mu$ = 0.7. The strength $V_{0}$ is included in the resulting normalization. We account coherent contributions from different combinations ($l^{\prime}I^{\prime}j^{\prime},lIj$) that allow transferred momenta $L$ = 1, 3, 5. The calculations are carried out with code DWUCK4\cite{28}. 

Figure 8 shows that the better description of the angular distribution is given by a component with $L$ = 1 including small corrections from other components. In addition to this calculation, the analysis in the framework of the MDM is performed (see Fig. 9). The shape of the angular distributions is approximated by combination of the Bessel functions $J_{L}$(x), where $L$ is the transferred angular momentum. 


\begin{figure}
\centerline{\includegraphics[width=6 cm]{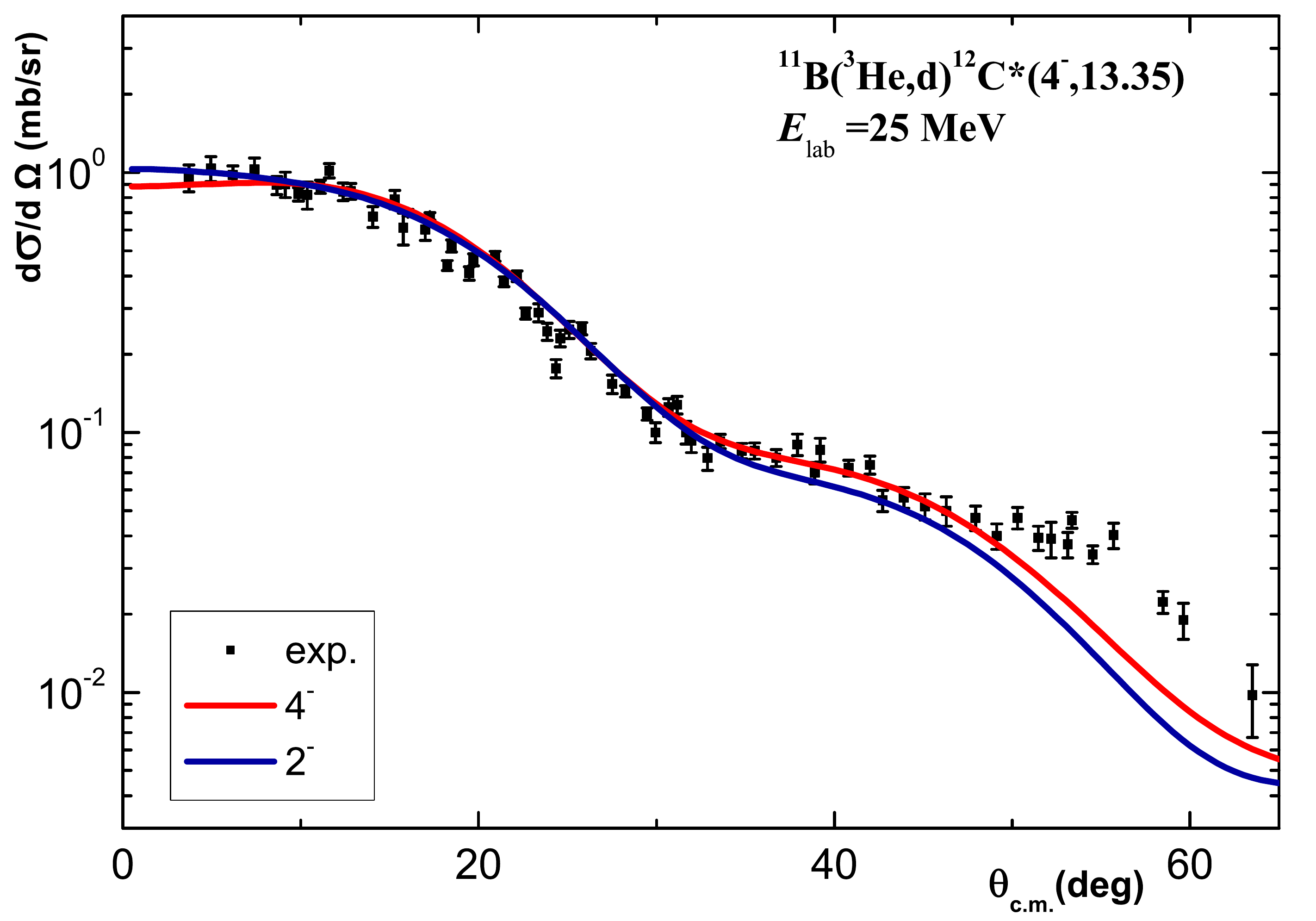}}
\caption{
The experimental differential cross-section of the $^{11}$B($^{3}$He,d)$^{12}$C reaction at $E_{lab}$ = 25 MeV with excitation of the 13.35 MeV state (black dots). The red and blue solid curves correspond to the DWBA calculations with the spin-parities of 4$^{-}$ and 2$^{-}$ ($L$=2), respectively.
}

\end{figure}

\begin{figure}
\centerline{\includegraphics[width=6 cm]{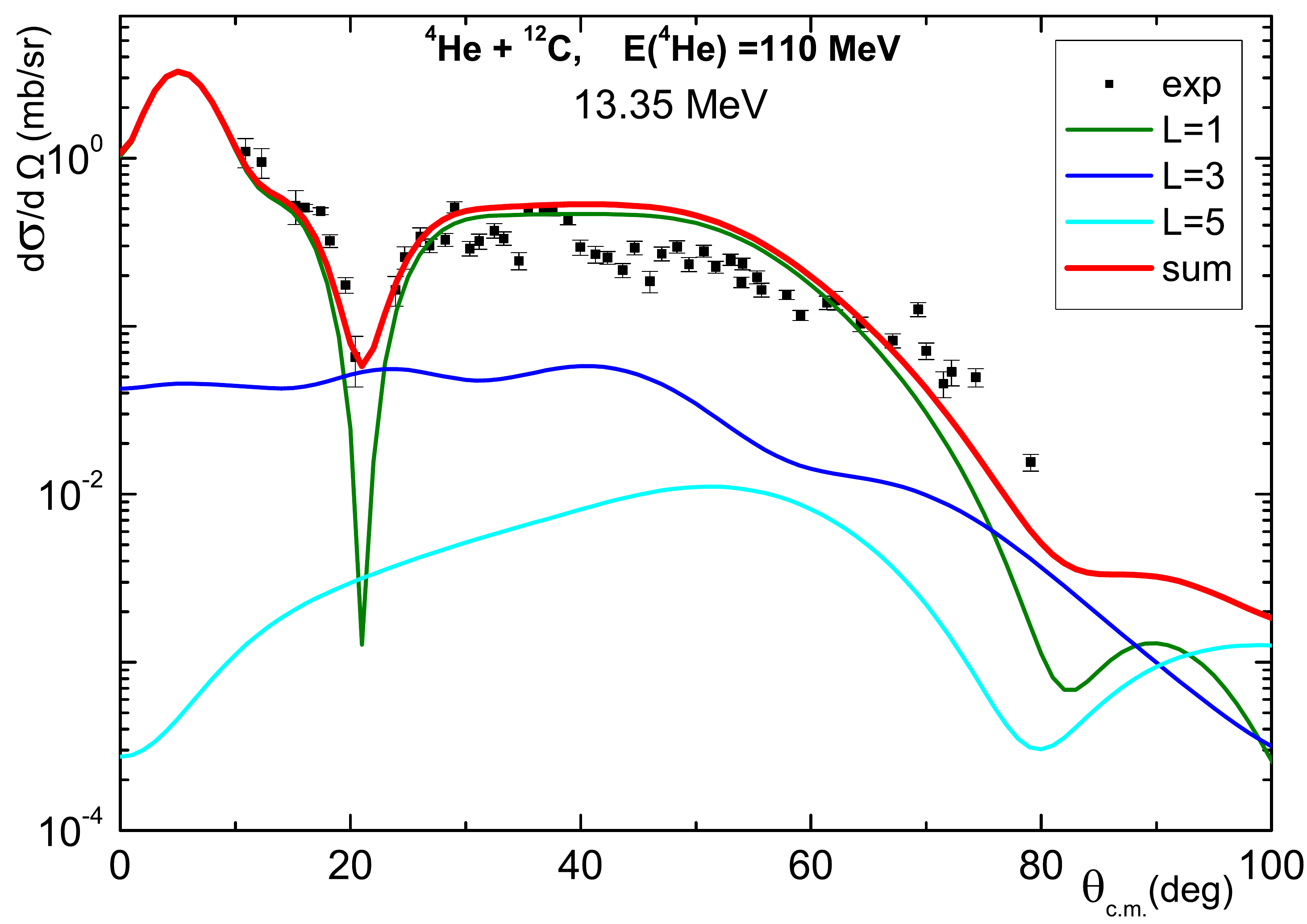}}
\caption{
The experimental differential cross section of the $^{4}$He + $^{12}$C inelastic scattering at $E_{lab}$ = 110 MeV with an excitation of the 13.35-MeV state (black dots). The green, blue, and cyan solid curves correspond to the DWBA calculations with $L$ = 1, 3, 5, respectively. The red solid curve depicts their coherent sum.
}
\end{figure}

\begin{figure}

\centerline{\includegraphics[width=6 cm]{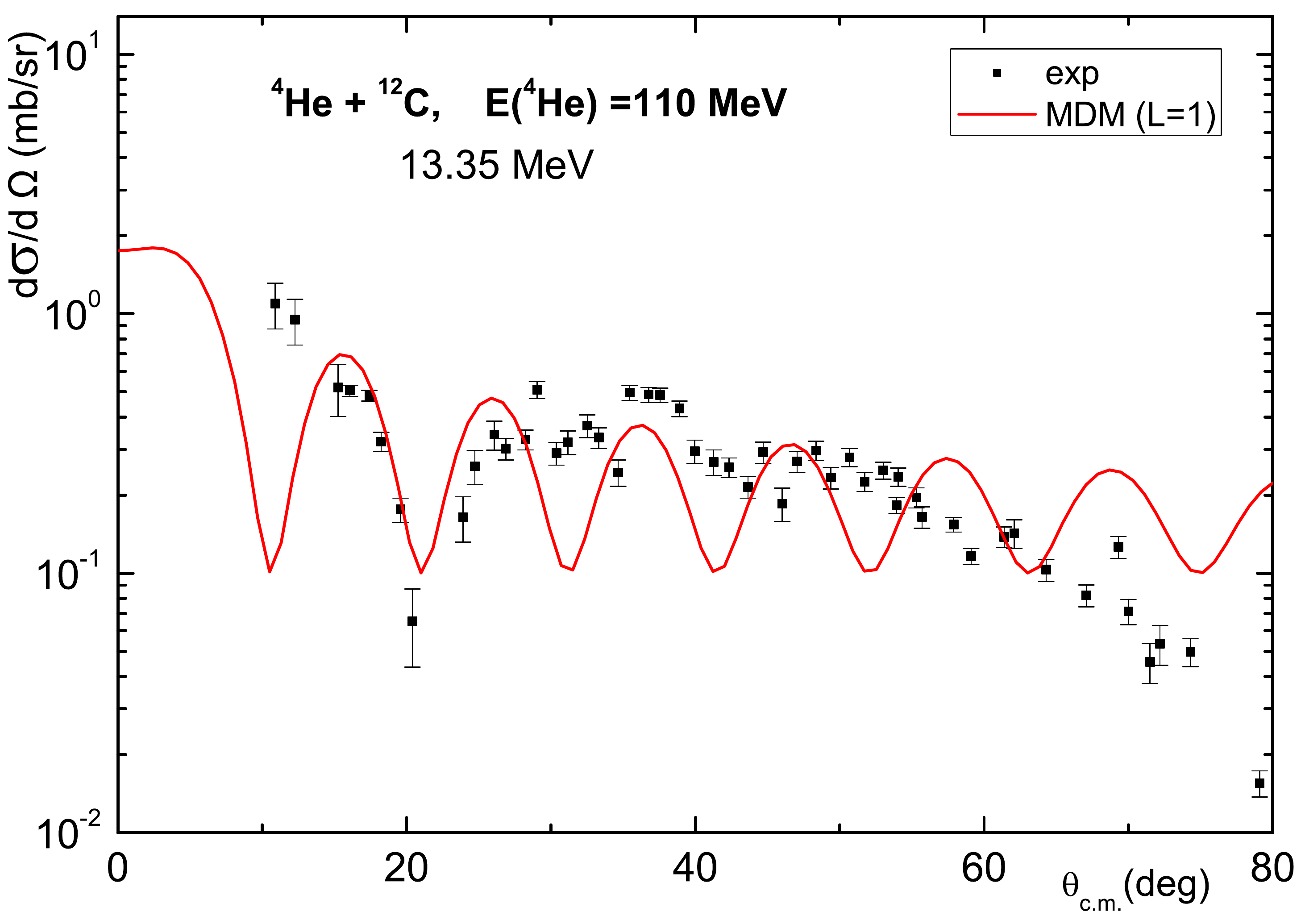}}
\caption{
The experimental differential cross section of the $^{4}$He + $^{12}$C inelastic scattering at $E_{lab}$ = 110 MeV with excitation of the 13.35-MeV state (black dots). The red solid curve corresponds to the Bessel function with $L$ = 1.
}
\end{figure}

As we can see in Fig. 9, the most probable transferred angular momentum also is $L$ = 1. The MDM analysis allowed us to estimate the rms radii of $^{12}$C in the excited state at 13.35 MeV, given a value of $R_{rms}$ $\approx$ 3 fm. This increased value is close to the radius of the 3$^{-}$ 9.64-MeV state of $^{12}$C ($R_{rms}$ = 2.88 $\pm$ 0.11)\cite{6}. Thus, if we assume that the 9.64-MeV and the 13.35-MeV states belong to a rotational band based on the 9.64-MeV state, then it is quite natural to assume that the spin-parity of the 13.35 MeV state should be 4$^{-}$. 

\subsection{The state at 22.4 MeV}

The differential cross section of the $^{11}$B($^{3}$He,d)$^{12}$C reaction with excitation of the 22.4-MeV state in $^{12}$C is presented in Fig. 10. The DWBA analysis of the data was carried out taking into account all possible spin-parities for this state. The best fit of the data was obtained for the spin-parities of 6$^{+}$ and 5$^{-}$. The calculated cross sections are also shown in Fig. 10 as green and blue curves. From our point of view, the fit of the data in the region of the main peak is better for the calculations corresponding to the spin-parity of 6$^{+}$. This is especially evident if we compare the experimental and theoretical differential cross sections in a linear scale (see the right panels in Fig. 10). 

As the state at 22.4 MeV was also observed in the inelastic scattering of $\alpha$-particles at $E_{lab}$ = 110 MeV\cite{14,23}, we can assign the isospin of $T$ = 0 to the 22.4-MeV state of $^{12}$C.

\begin{figure} [th]
\centerline{\includegraphics[width=10 cm]{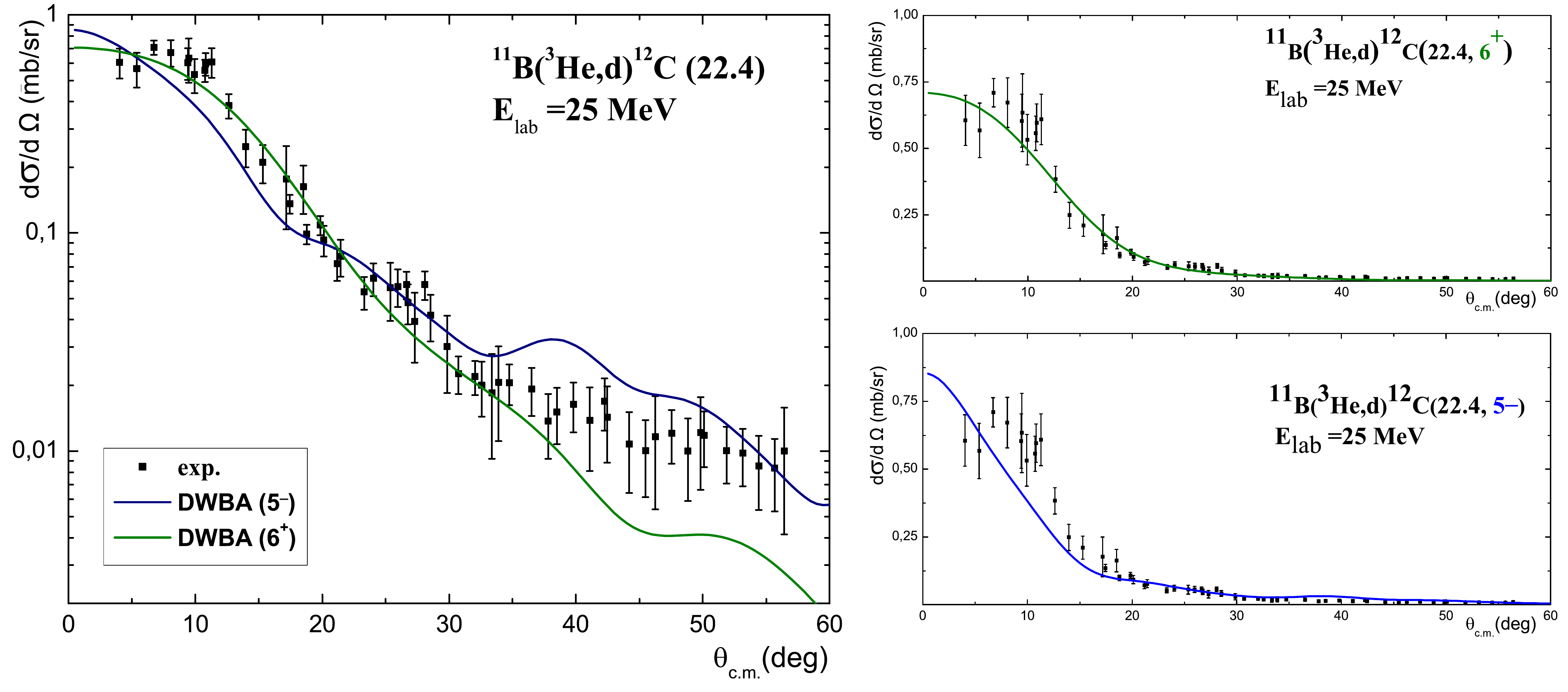}}
\caption{
Left panel: the experimental differential cross section of the $^{11}$B($^{3}$He,d)$^{12}$C reaction for the 22.4-MeV state (black dots) in comparison with the DWBA calculations with spin-parities of 5$^{-}$ (blue curve) and 6$^{+}$ (green curve). Right panel: a comparison of the data with the DWBA calculations in a linear scale.
}
\end{figure}

If we accept that the spin-parity of the 22.4-MeV state is of 6$^{+}$ and the isospin is $T$ = 0 then it corresponds to the rotational trajectory $J (J + 1)$ of the Hoyle band, which also includes the 2$^{+}$ state near 9.9 MeV\cite{9, 10, 11} and the 4$^{+}$ state at 13.75 MeV\cite{20,23}.

\subsection{New members of the rotational band in $^{12}$C}

To summarize the analysis of the received data let us present an updated systematics of the states in $^{12}$C grouping them to the rotational bands shown in Fig. 11. The well-established $K^{\pi}$ = 0$^{+}$ ground-state rotational band includes 2$^{+}$ and 4$^{+}$ states at 4.44, and 14.08 MeV (a black line in Fig. 11). The $K^{\pi}$ = 3$^{-}$ rotational band proposed in Ref.\cite{15} contains the negative-parity states 3$^{-}$ and  4$^{-}$ states at 9.64 and 13.35 MeV (a blue line in Fig. 11). The rotational band based on the Hoyle state can include the 0$^{+}$, 2$^{+}$, 4$^{+}$, and 6$^{+}$ states at 7.65, 9.9 (9.75\cite{11} or 10.13\cite{9}), 13.75 and 22.4 MeV (shown in Fig. 11 by red line).

\begin{figure} [th]
\centerline{\includegraphics[width=8 cm]{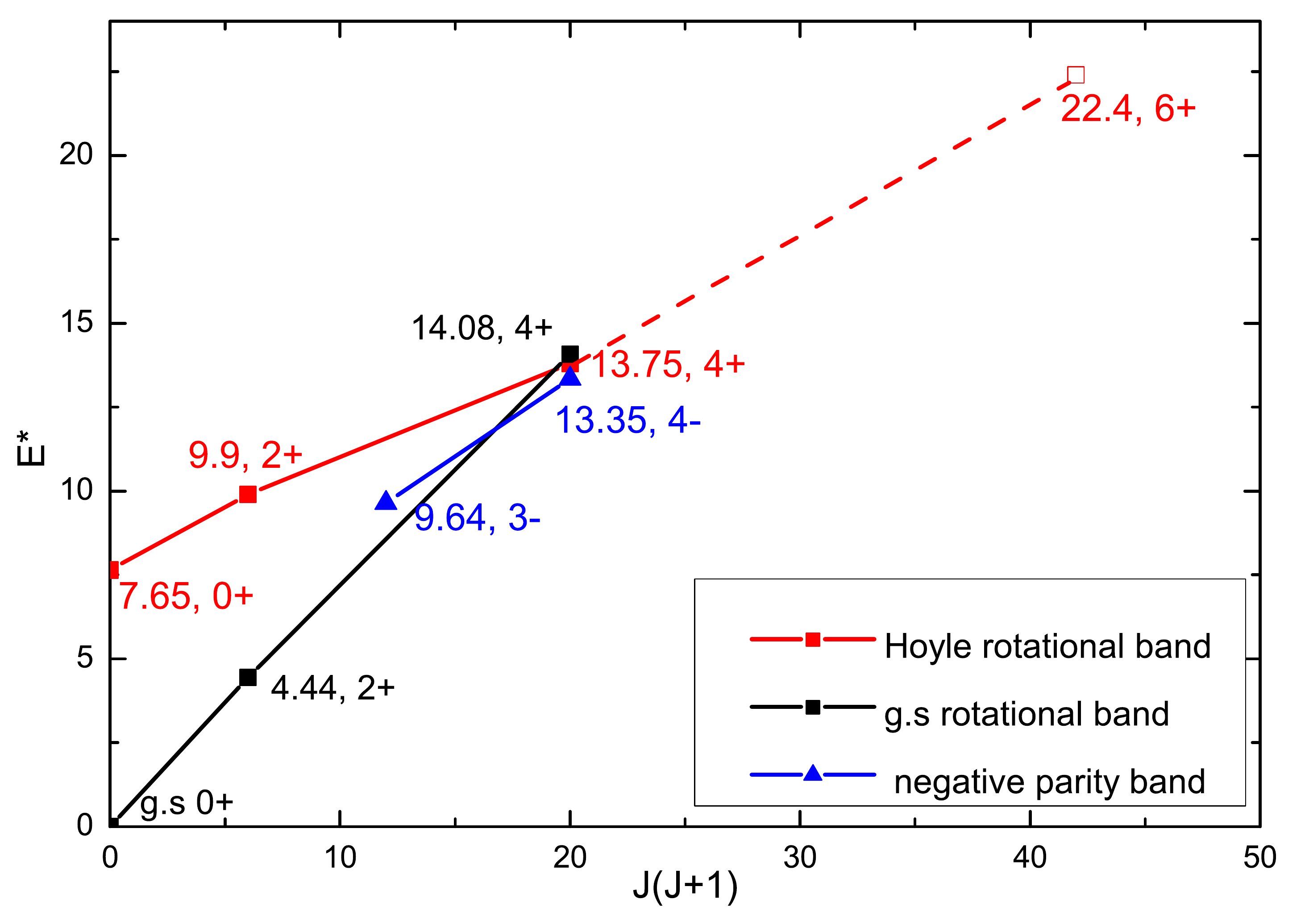}}
\caption{
Rotational bands in $^{12}$C. The g.s band (black line), the negative-parity band (blue line), the exotic rotational band based on the Hoyle state (red line). A proposed missing member of this band, 6$^{+}$ at 22.4 MeV is marked by open point.
}
\end{figure}

Recently the observation of the state at 22.5 MeV with a tentative assignment, 5$^{-}$, was announced\cite{15}. Unfortunately, nothing is stated in Ref.\cite{15} regarding the isospin of the new state. However, the new data table assigns $T$ = 1 in the energy region 22.4 MeV\cite{29}. The authors assumed that it is a good candidate to complete the negative-parity rotational band. 

The matter is further complicated by the fact that in this high-excitation region of the spectrum, the majority of states overlap and therefore their separation and characterization is not straightforward. It is conceivable that in the excitation region of 22.4-22.5 MeV, there are two overlapping states with spin-parities of 5$^{-}$ and 6$^{+}$. The case would be therefore similar to the 21.6-MeV state having both 2$^{+}$ and 3$^{-}$ as accepted spin-parity values.

\section{Conclusion}

The $^{11}$B($^{3}$He,d)$^{12}$C reaction was chosen to study proton transfer at $E_{lab}$ = 25 MeV. Deuteron angular distributions for the g.s. and eight excited states at $E_{x}$ = 4.44, 7.65, 9.64, 13.35, 16.57, 20.98, 21.6 and 22.4 MeV of $^{12}$C were measured in the c.m. angular range 4$^{\circ}$-60$^{\circ}$. In addition, the data on inelastic scattering of $\alpha$-particles on $^{12}$C at 110 MeV with excitation of the 13.35 MeV state are presented for the first time.

The generally accepted spin-parity of 2$^{-}$ and the isospin of $T$ = 1 for the state at 16.57 MeV, and the spin-parity of 2$^{+}$ and 3$^{-}$, and the isospin of $T$ = 0 for the state at 21.6 MeV were confirmed. For the first time, the spin-parity, 3$^{-}$, and the isospin $T$ = 0 for the state at 20.98 MeV were determined.

The DWBA analysis of the differential cross section of the $^{11}$B($^{3}$He,d)$^{12}$C reaction with excitation of the 13.35 state reveals that the spin-parity assignments of 4$^{-}$ and 2$^{-}$ have equal probabilities. Nevertheless, an additional analysis of the inelastic scattering of $\alpha$-particles on $^{12}$C at 110 MeV and an estimate of the radius of this state by the MDM provide additional arguments for a tentative assignment of the spin-parity 4$^{-}$ and the isospin $T$ = 0 for this state confirming the statements in Ref.\cite{18, 19} and a proposal\cite{5} of a formation of the negative parity band in $^{12}$C by the states 9.64 MeV, 3$^{-}$ and 13.35 MeV, 4$^{-}$.

The state at 22.4 MeV has attracted particular interest in recent years\cite{5,18,19}. It was proposed as a candidate for the third position in the negative parity band, providing it has a spin-parity of 5$^{-}$. Our analysis has shown this state could rather be assigned spin-parity of 6$^{+}$ and isospin $T$ = 0. We believe that, most likely, it is the fourth member of the rotational band based on the Hoyle state. At the same time, it cannot be excluded that in the excitation region of 22.4 MeV there are two unresolved states with spin-parities of 5$^{-}$ and 6$^{+}$.

\section*{Acknowledgements}

The work was partly supported by the Russian Science Foundation (project no. 18-12-00312) and a mobility grant from the Academy of Finland. We are very grateful to the staff of the Accelerator Laboratory for excellent working atmosphere and maintaining high-quality beams throughout the measurements.

\end{document}